\documentclass{article}

\usepackage[preprint, nonatbib]{neurips_2020}

\usepackage[utf8]{inputenc} 
\usepackage[T1]{fontenc}    
\usepackage{hyperref}       
\usepackage{url}            
\usepackage{booktabs}       
\usepackage{amsfonts}       
\usepackage{nicefrac}       
\usepackage{microtype}      
\usepackage{xcolor}
\usepackage{graphicx}
\usepackage{subcaption}
\usepackage{verbatim}
\usepackage{float}

\title{PersGNN: Applying Topological Data Analysis and Geometric Deep Learning to Structure-Based Protein Function Prediction}

\author{
  Nicolas Swenson\thanks{Equal contribution. \texttt{\{nswenson, akrishnapriyan\}@lbl.gov}}
  \And
  Aditi S. Krishnapriyan\footnotemark[1] \\
  \AND
  Aydin Buluc \\
  \And
  Dmitriy Morozov
  \And
  Katherine Yelick
  \AND
  \\
  Lawrence Berkeley National Laboratory \&\\
  Department of Electrical Engineering and Computer Science,\\ University of California, Berkeley\\ 
  Berkeley, CA, 94720 \\
}

\begin{document}

\maketitle

\begin{abstract}

Understanding protein structure-function relationships is a key challenge in computational biology, with applications across the biotechnology and pharmaceutical industries. While it is known that protein structure directly impacts protein function, many functional prediction tasks use only protein sequence. In this work, we isolate protein structure to make functional annotations for proteins in the Protein Data Bank in order to study the expressiveness of different structure-based prediction schemes. We present PersGNN---an end-to-end trainable deep learning model that combines graph representation learning with topological data analysis to capture a complex set of both local and global structural features. While variations of these techniques have been successfully applied to proteins before, we demonstrate that our hybridized approach, PersGNN, outperforms either method on its own as well as a baseline neural network that learns from the same information. PersGNN achieves a $9.3\%$ boost in area under the precision recall curve (AUPR) compared to the best individual model, as well as high F1 scores across different gene ontology categories, indicating the transferability of this approach.

\end{abstract}

\section{Introduction}

Predicting protein function from its raw amino acid sequence is a long standing challenge in computational biology. This is an ideal setting for computational methods because experimental characterization of proteins is costly and time-consuming, and there is an abundance of protein sequence data. Public databases such as Uniprot contain over 100 million protein sequences \cite{uniprot2019uniprot}. The Critical Assessment of protein Function Annotation (CAFA), a recurring challenge that benchmarks different computational approaches, has shown that machine learning and statistical algorithms outperform alignment-based techniques, such as BLAST, or homology transfer methods \cite{zhou2019cafa, zhu2019cafa}. 

 However, proteins are not just sequences of amino acids: they fold into complex three-dimensional motifs, which directly impact a protein's function \cite{berg2002biochem}. Recent work has shown that using protein sequences together with its three-dimensional structural information can lead to better functional predictions, as well as provide scientists with a way to identify the functionally-active areas of a protein \cite{gligorijevic2020gcn}. This work is made possible by advances in structural biology, such as X-ray crystallography, which have allowed for the structure of many proteins to be determined. For example, the Protein Data Bank (PDB) contains structural information for over 100,000 proteins and other biological molecules \cite{berman2003pdb}.
 
Given the success of these new structure-aware machine learning techniques, in this work, we seek to maximize the usefulness of structure for functional characterization. Since protein structure determination can be laborious and expensive, optimal utility of these structures is of particular importance. We introduce PersGNN, a method that combines topological data analysis (specifically, persistent homology) and graph neural networks to create a more nuanced representation of the protein structure. To the best of our knowledge, this is the first time that these approaches have been combined in this way. Since determining the protein structure via experimental methods is challenging, maximizing the utility of the structural data is key. We show that our hybrid model learns more from structure than either model on its own, and also outperforms a standard neural network that is given the same information. In future work, protein sequence models can also be incorporated in order to create even better functional annotation models.

\section{Methods}

\subsection{Graph Neural Networks}
\label{sec:gnn}

Graph Neural Networks (GNN), which can be motivated through spectral graph convolutions, are a popular form of graph representation learning that provide a framework for extending traditional deep learning techniques to non-Euclidean, graphical data \cite{kipf2017semisupervised, bruna2014spectral}. GNNs learn context-aware node embeddings through successive rounds of local neighborhood aggregation (or "message passing") in a graph. These node embeddings can be combined to create global representations for entire graphs. Previous work has used 3D Convolutional Neural Networks (3D CNNs) to extract useful information from protein structure \cite{jimenez2017cnn, amidi2018cnn}. However, this technique is both memory and computationally inefficient because proteins are sparse in 3D space and 3D CNNs will perform many convolutions over empty space. More recent work proposed to overcome some of these inefficiencies with a GNN \cite{gligorijevic2020gcn}. In this method, 3D protein structures are represented by contact maps, which threshold the pairwise distances between the alpha-carbons of each residue in the protein. These contact maps are then fed into the GNN, which embeds structural information that can used to make function predictions. 

Our GNN model architecture is inspired by the one of \textit{ Gligorijevic et al.}, which uses the graph convolutional layers from \textit{Kipf and Welling}, along with language modeling \cite{gligorijevic2020gcn, kipf2017semisupervised}. In contrast, our work focuses on learning from structure. Thus, we omit the language modeling component and instead opt for a simple one-hot representation of amino acids. Other modeling differences, such as model depth and learning rate, were optimized during the validation stage.

\subsection{Persistence Network}
\label{sec:persnet}

We use persistent homology, an area of topological data analysis, to construct a topological representation of the protein structure. In this section, we briefly describe the approach and refer the interested reader to a more thorough survey \cite{edelsbrunner2008persistent}.

We start with the 3D atomic coordinates of a protein structure and take the union of balls around the atoms. We vary the radii of these balls, and track how the topology of this union changes: these are represented as alpha shapes \cite{edelsbrunner1994three}. By doing this and sweeping across all radii, we get an increasing sequence of nested alpha shapes called the \textit{filtration.} By following the changes in the sequence, one can keep track of the appearances and disappearances of topological features in the filtration.

We record the pairs of radii when such topological features appear and disappear, where the appearance of a new feature is the \textit{birth} and the disappearance of this same feature (such as when it gets merged with another feature) is the \textit{death}. The difference between the \textit{birth-death pairs}, i.e. \textit{death - birth}, is called the \textit{persistence} of the feature. The larger the \textit{persistence}, the more prominent the topological feature. The full set of all \textit{birth-death pairs} is called a \textit{persistence diagram}. In this work, we specifically look at the 1-dimensional and 2-dimensional persistence diagrams of each protein structure, where 1-dimensional features correspond to channels (``loops'') and 2-dimensional features correspond to voids (``cavities'') in the protein structure.

Feeding persistence diagrams, in this case the topological summaries of all the channels and voids in a protein structure, into a neural network architecture requires a transformation of the diagram points. An approach to do this was first proposed by Hofer et al. \cite{hofer2019learning}. Expanding on this, Carri\`{e}re et al. \cite{carriere2020perslay} proposed a layer for neural network architectures based on the DeepSet architecture \cite{zaheer2017deep} to encode a vector representation of the persistence diagram through point-wise transformations. We implement a modification of this approach to process both 1D and 2D persistence diagrams of the protein structure, which we call ``PersNet'' in this work.

The problem of describing protein shape was one of the early motivations behind persistent homology \cite{edelsbrunner2000topological}. Persistent homology has also been used previously to construct topological representations to describe protein structures for protein classification problems \cite{cang2015topological, dey2018protein}. However, these approaches turn topological information into feature vectors and rely on the construction of ad hoc handcrafted summaries. More recently, vectorizing persistence diagrams in ways that require minimal processing and retain more of the information, such as to persistence images \cite{adams2017persistence}, have been used in machine learning applications for scientific domains \cite{rieck2020uncovering, krishnapriyan2020topological, krishnapriyan2020persistent, hong2020classifying}. However, these vectorized representations are static --- in contrast, the approach outlined here processes the topological information through an automatically differentiable layer, thereby continuing to learn the representation over the training process.

\subsection{PersGNN: Hybrid network}

\begin{figure}
    \centering
    \includegraphics[width=.9\linewidth]{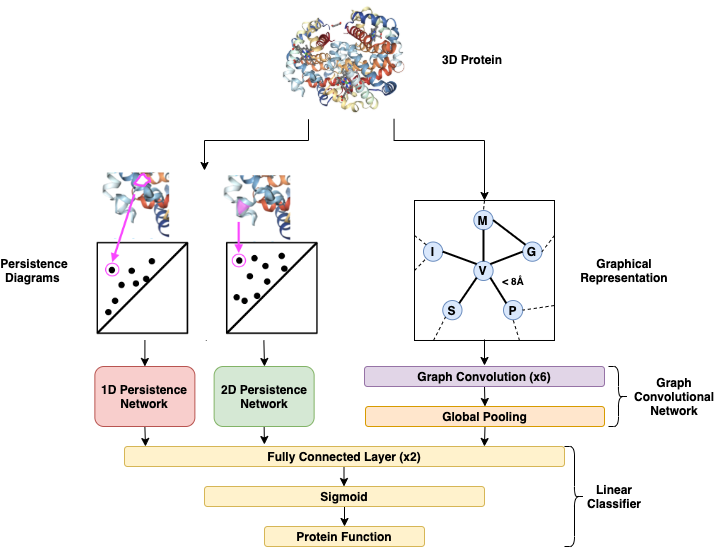}
        \caption{\textbf{PersGNN model architecture.} Starting from a protein's 3D structure, we compute 1D and 2D persistence diagrams (summaries of all of the topological features, i.e. channels and voids, in a protein) and $C_\alpha-C_\alpha$ contact maps. The persistence networks (PersNets) compute a vectorized representation of each persistence diagram. The graph neural network (GNN) combines the contact map with one-hot encoded amino acid labels to learn a separate vectorization for the whole structure. These representations are concatenated together and fed into a two-layer multi-layer perceptron (MLP) to predict molecular function (MF) gene ontology (GO) terms. For the individual models (Sections~\ref{sec:gnn} and ~\ref{sec:persnet}), we use only the GNN and PersNet pipelines respectively.} 
    \label{fig:model}
\end{figure}

Using a GNN in place of a 3D CNN significantly reduces computation time and memory requirements of learning on protein structures, but also requires representing the protein structure with a contact map. This simplification is necessary for the GNN to work but can also lead to a loss of key structural information. To overcome this, we propose incorporating persistent homology into the learning process by converting the 3D atomic coordinates of the protein structure to persistence diagrams and processing this information through a layer in the neural network architecture, as described in~\ref{sec:persnet}. We note that previous work has used persistent homology together with 3D CNNs to classify protein folds \cite{hong2020classifying}; however, as mentioned earlier, they used a static vectorization of the persistence diagrams of the protein structure rather than a neural network processing layer that learns the representation through training and did not use GNNs --- both things that we found very important in our approach to achieving high accuracy. There has also been previous work combining persistent homology and GNNs \cite{zhao2020persistence}, but here they incorporated persistent homology into the graph neural network architecture itself by using persistent homology to reweight messages passed between the graph nodes during convolutions. Our method allows us to build on the successes of applying GNNs to protein data (and capitalize on their computational efficiency) while also adding additional structural features that are potentially not captured by a protein's contact map, such as more global topological information. 

Our hybrid model, shown in Figure~\ref{fig:model}, concatenates the output from the GNN with the output from the PersNet models that process both the 1D and 2D topological features (channels and voids respectively). This hybrid representation is then passed into a Multi-Layer Perceptron (MLP) to perform the classification task. The architecture of the GNN and PersNet models is the same as the ones described in Sections~\ref{sec:gnn} and ~\ref{sec:persnet}. PersGNN is trained end-to-end on a protein's contact maps, amino acid labels, and persistence diagrams and all sub-components are trained simultaneously. This enables PersGNN to learn a complex representation of protein structure efficiently and with minimal manual processing.

\subsection{Baseline}

In addition to benchmarking our hybrid model against its component models, we trained a simple three-layer MLP to create a more complete picture of how these models are learning from structure. This is distinct from many assessments of protein functional annotations, such as CAFA, which use the alignment-based BLAST as a baseline. Since this investigation is focused on extracting information from protein structure, rather than its amino acid sequence, we decided not to use the sequence-only BLAST (or any other sequence-based methods) as a baseline. The baseline model is given all the same information as the GNN: for each residue, the model is given the amino acid label and the number of residue ``contacts,'' which are computed from the same contact maps that are fed into the GNN. 

\subsection{Data and Model Parameters}

Our dataset consists of 33,007 proteins taken from PDB that belong to the Arabidopsis, Celegans, E.coli, Fly, Human, Mouse, Yeast and Zebrafish species. This is randomly split into train, validation and test sets at a $75\%$, $12.5\%$, $12.5\%$ ratio. Random splitting can sometimes be overly optimistic and not representative of realistic challenges \cite{sheridan2013split}. Previous work, for example, has split their datasets to minimize $\%$ sequence identity between training and tests set, which we leave for future work.

For labels, we represent protein function with molecular function (MF) gene ontology (GO) terms. For each term in PDB, we obtains its corresponding MF GO terms using the Structure Integration with Function, Taxonomy and Sequence (SIFTS) database \cite{velankar2018sifts, dana2018sifts}. Consistent with previous studies, we filter out GO terms with less than 25 representative proteins. After the filtration, our label space is made up of 730 MF GO terms. Each protein can also have multiple labels, making this a multi-label multi-classification problem.

Each protein entry in PDB is described by a list of amino acid residues and their 3D atomic coordinates. For the GNN and MLP models, the 3D atomic coordinates are converted to $C_\alpha-C_\alpha$ contact maps. To construct these contact maps, we compute pairwise distances between the $\alpha$-carbon of each residue and denote a "contact" between residues if their $\alpha$-carbons are within eight angstroms. The GNN and MLP models also take a one-hot encoded representation of amino acids, which the GNN uses as node labels. We also construct 1D and 2D persistence diagrams from the 3D atomic coordinates of each protein structure, which are then processed through the persistence network layers. 

All models are trained using a weighted binary cross-entropy loss function \cite{gligorijevic2020gcn}, the ADAM optimizer \cite{kingma2017adam}, for 300 epochs and with a learning rate of $10^{-5}$. To normalize our models, we employ weight normalization and dropout regularization (p = 0.1) throughout \cite{salimans2016weight, srivasta2014dropout}. For each method, we create an ensemble by independently training 10 models and taking the average bit-score (per GO term) across the 10 models at evaluation time.

\section{Results and Discussion}

\begin{figure}
\begin{minipage}{0.6\textwidth}
    \centering
    \includegraphics[width=\linewidth]{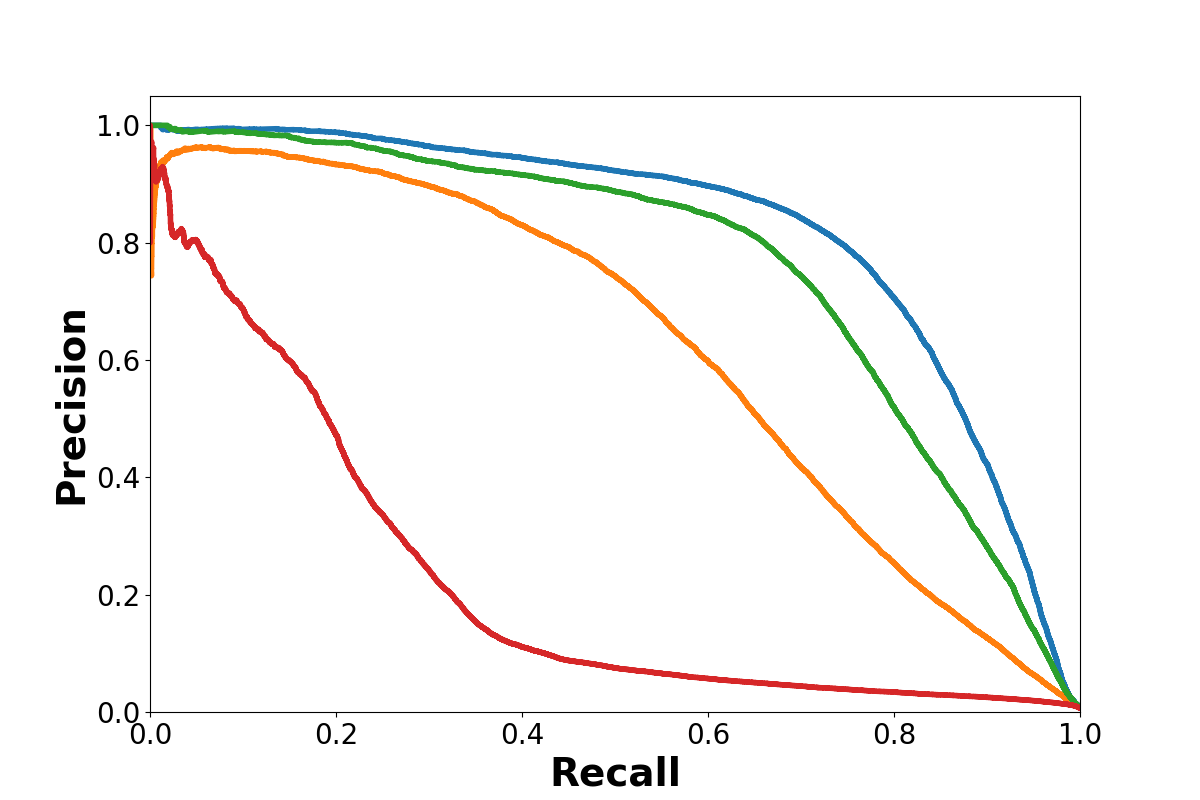} 
\end{minipage}
\begin{minipage}{0.19\textwidth}
\centering
\begin{tabular}{ c | c }
Model & AUPR \\
\hline
\textcolor{blue}{\textbf{---}} PersGNN & \textbf{0.82} \\
\textcolor{green}{\textbf{---}} GNN & 0.75 \\
\textcolor{orange}{\textbf{---}} PersNet & 0.63 \\
\textcolor{red}{\textbf{---}} MLP (Baseline) & 0.22 \\
\end{tabular}
\end{minipage}
\caption{\textbf{Precision-recall curves for predicting molecular function (MF) gene ontology (GO) terms.} Precision-Recall curves for four models: the "baseline" model (a multi-layer perceptron) trained on protein contact maps, a persistence network (PersNet) trained on the persistence diagrams created from the 3D atomic coordinates of each protein, a graph neural network (GNN) trained on protein contact maps, and PersGNN, our model that combines PersNet and the GNN. For each method, we independently train ten models and ensemble them by computing the average probability score for each class. Area under the precision-recall curve (AUPR) scores on the test set for each method are shown to the right, where higher scores indicate better accuracy. Our model, PersGNN, achieves the highest AUPR score.}
\label{fig:pr}
\end{figure}

Our hybrid method, PersGNN, outperforms both GNN and PersNet on their own, and significantly outperforms the baseline MLP that is given the same information as the GNN. The performance of each method, measured in area under the precision-recall curve (AUPR) for molecular function (MF) gene ontology (GO) terms is shown in Figure \ref{fig:pr}. PersGNN has an AUPR score $9.3\%$ higher than the GNN, the next best model. To focus our study on learning from protein structure, we have not included highly expressive sequence models, such as BLAST or 1D CNNs, nor did we use language models to compute amino acid embeddings. The GNN, however, can learn to embed amino acids through a residue's local neighborhood in the graph structure. PersNet is able to capture further topological information through a protein's 1D and 2D persistence diagrams. When combined, the GNN and PersNet capture complementary information as indicated by the higher AUPR score, thus creating a more complete representation of the protein structure. 

\begin{figure}
\centering
\includegraphics[width=\linewidth]{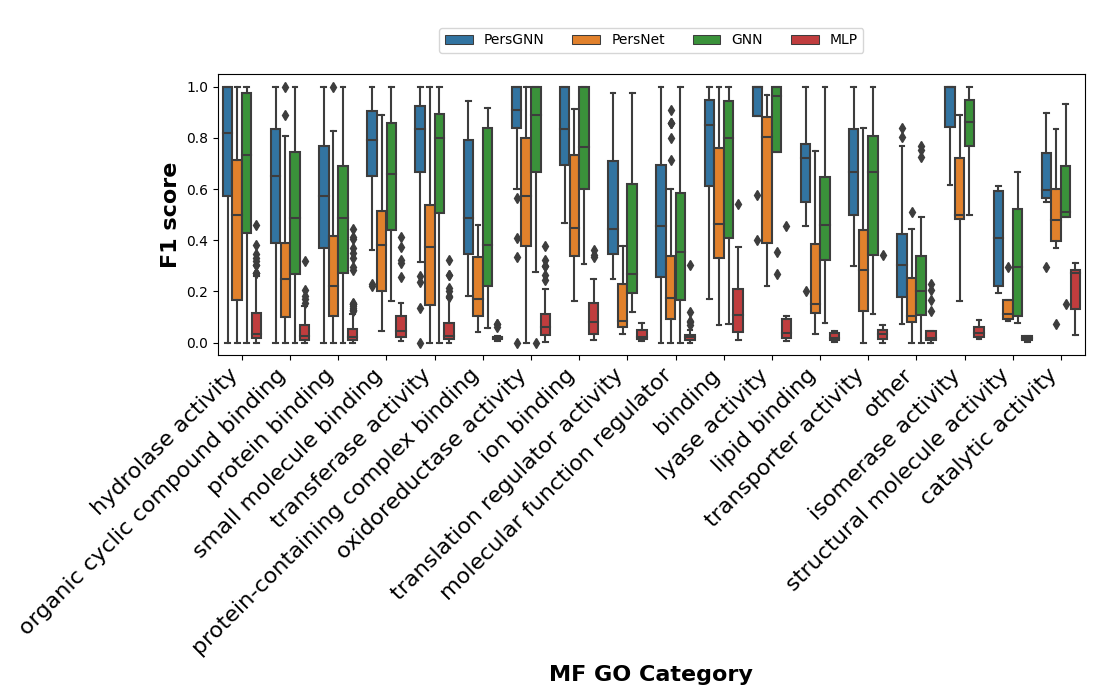}
\caption{\textbf{Model performance by GO term type.} Boxplot showing average F1 scores (higher is better) aggregated over umbrella GO term categories for all four models. PersGNN achieves the highest average F1 scores across all MF categories, showing the transferability of this approach. This indicates that the topological representation from persistent homology is learning complementary structural information to the GNN, such as the ability to capture more global information. As a result, PersGNN is able to learn a more nuanced representation of structure that is important for making accurate functional annotations.}
\label{fig:go_type}
\end{figure}

In Figure~\ref{fig:go_type}, we compute average F1 scores aggregated over different GO categories, which are grouped at various levels of the MF GO hierarchy. F1 scores are a measure of the model accuracy, calculated from the precision and recall, where higher F1 scores indicate that the model was able to successfully classify more proteins. As we see in Figure~\ref{fig:go_type}, the PersGNN model has consistently high F1 scores across GO categories, and performs better than the other methods on every GO category and almost every individual GO term.

\begin{figure}
\begin{subfigure}{.5\textwidth}
    \centering
    \includegraphics[width=\linewidth]{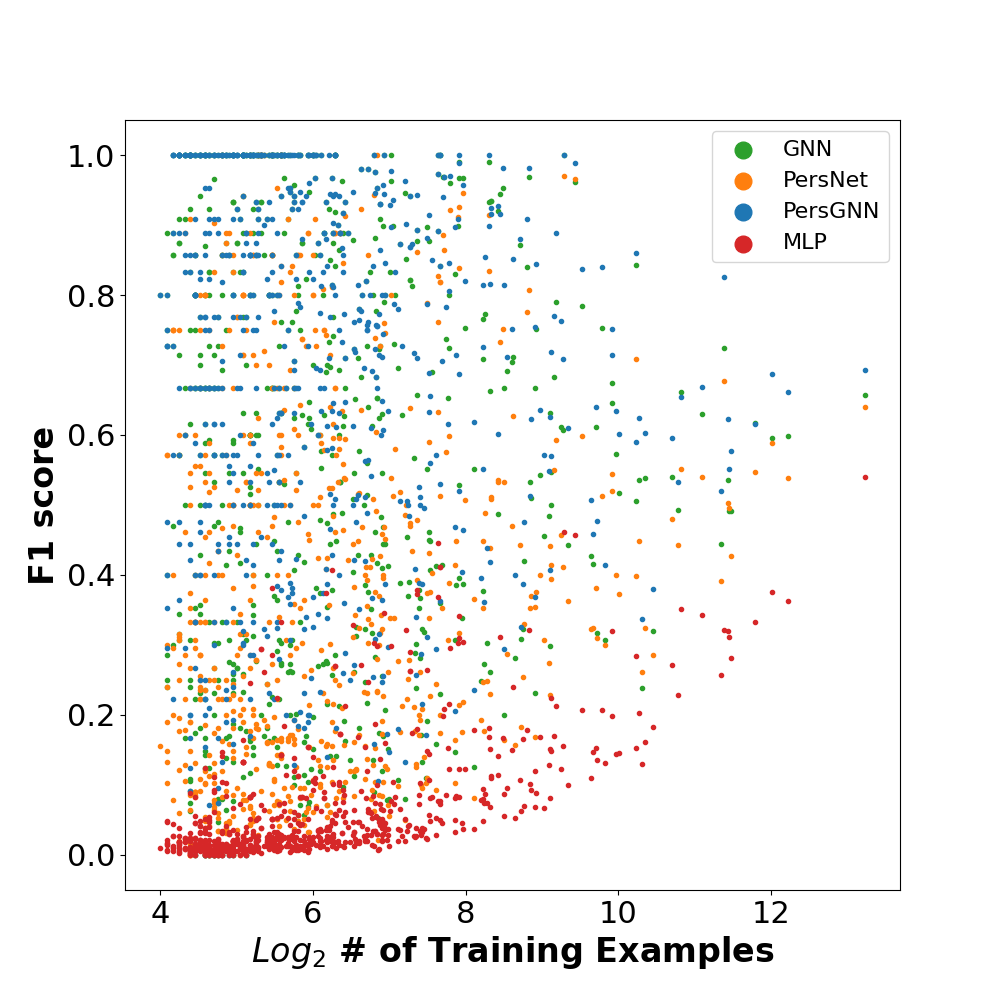}
\end{subfigure}
\begin{subfigure}{.5\textwidth}
    \centering
    \includegraphics[width=\linewidth]{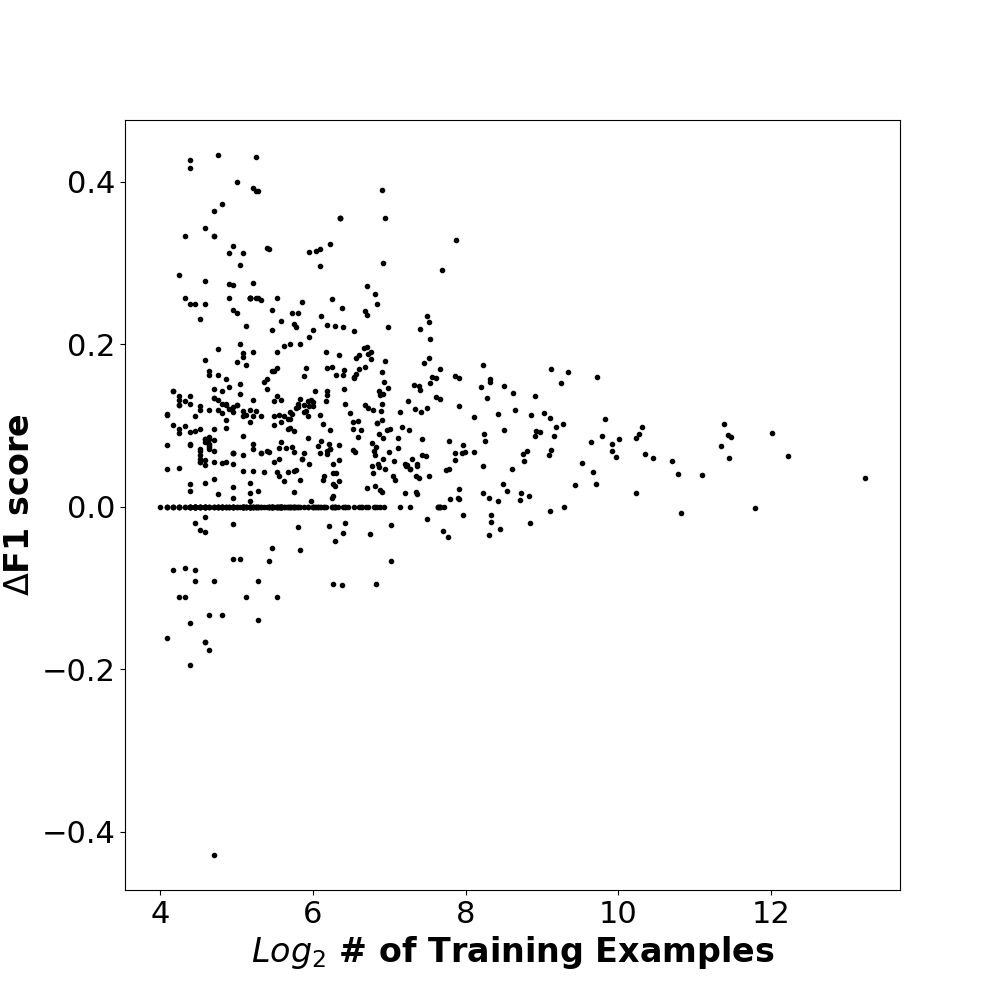}
\end{subfigure}
\caption{\textbf{Effect of number of training examples on model performance.} F1 scores are plotted by MF GO term for all four models, where higher F1 scores indicates more accuracy (left) and the change in F1 score per GO term from PersGNN to GNN, where positive $\Delta$ means PersGNN scored higher (right). PersGNN outperforms other models on GO terms with fewer training examples. This suggests that PersGNN is constructing more meaningful structural features, enabling it to learn from fewer training examples.}
\label{fig:num_examples}
\end{figure}

Figure \ref{fig:num_examples} shows the effects of training set size (the number of times each GO term appears in the training dataset) against model accuracy, again represented via F1 scores. As we see, PersGNN achieves high F1 scores even on GO terms with fewer training examples, while other models like the MLP perform poorly in this regime. The ability of PersGNN to make accurate predictions even with a low training set size is optimistic, as it is indicates the model is making good use of the protein structure information. Moreover, while there are millions of raw amino acid sequences, there are far fewer available protein structures, meaning achieving high model accuracy with lower amounts of data is especially important here.

Our method, PersGNN, more accurately predicts MF GO terms compared to other structure-based methods, including across different categories and with fewer examples. This motivates a further investigation into its performance. Future work in this area should study PersGNN's performance on all three GO term categories (Biological Process and Cellular Component). In addition, it is known that random splits are often too optimistic, so future work should also evaluate performance on curated splits. We decided to limit the focus of this study to generating representations of protein structure and understanding protein structure-function relationships, despite the limited availability of high quality structural data. Other work has shown that structure-based methods trained on high quality structural data can still perform well when tested with predicted structures \cite{gligorijevic2020gcn}. With the continued advances in structure prediction methods (both De Novo and Ab Initio), this trend is likely to persist. As our model achieved high accuracy for protein function predictions using only the structural information of a protein structure (namely only through the 3D atomic coordinates), this confirms the value of incorporating structure into predictive models. Moreover, this approach requires minimal processing of the input data, making it a transferable across different datasets. 

\section*{Broader Impact}

Due the challenges of experimentally characterizing a protein's function, predicting protein function from sequence and structure is an ongoing challenge in the bioinformatics community (as evidenced by the recurring CAFA challenge). The ability to quickly and accurately annotate proteins using computational methods will have a great impact in metagenomics, drug-discovery, and many other biological applications. Many of the efforts to predict protein function rely on sequence alone, even though it is known that structure directly influences function. This is because sequences are much more abundant, due to rapid improvements in sequencing technology. As more and more protein structures become available, it will be important to incorporate structure into the function prediction pipelines.

Here, we present a new method that combines complementary representations of structure to improve functional annotations, achieving high accuracy by using only the protein structure. This is not a tool that is intended to be used in isolation; rather, it demonstrates that structural information and specifically persistent homology and GNNs should be part of the toolset used for analyzing proteins.  Combined with sequence information, we believe this will prove to be a powerful hybrid method for functional annotation. 

\begin{ack}


The authors would like to thank Jiali Chen, Jude Fernandes, Nick Bhattacharya and Andrew Tritt for their help and guidance. 

This work was supported by the U.S. Department of Energy under Contract Number DE-AC02-05CH11231 at Lawrence Berkeley National Laboratory. A.S.K. is an Alvarez Fellow in the Computational Research Division at LBNL. This research used resources of the National Energy Research Scientific Computing Center, which is supported by the Office of Science of the U.S. Department of Energy under Contract No. DE-AC02-05CH11231. This research was also supported by the Exascale Computing Project (17-SC-20-SC), a collaborative effort of the U.S. Department of Energy Office of Science and the National Nuclear Security Administration. The authors declare they have no competing financial interests.

\end{ack}

\bibliographystyle{unsrt}
\bibliography{main}

\end{document}